\documentstyle[11pt,paspconf,graphics]{article}

\begin{document}

\title{Single CO Peak in the Center of the Double-barred Galaxy NGC\,5850}

\author{S. Leon and  F. Combes}
\affil{DEMIRM, Observatoire de Paris, F-75014 Paris, France}

\author{D. Friedli}
\affil{Observatoire de Gen\`eve, CH-1290 Sauverny, Switzerland}








\section{Introduction}
NGC\,5850 is a prototype of double-barred galaxy (Friedli et al.
1996) classified as SBb(sr) I-II (Higdon et al. 1998; Prieto et al.
1997). This kind of system is primordial to understand the physical
mechanism responsible for feeding galaxy nuclei and boost the star
formation rate. Shlosman et al. (1989) proposed that the nuclear bar
would produce the inwards inflow of the molecular gas trapped on the
ILR of the primary bar through the nuclear bar resonances.  Higdon et
al. (1998) emphasized that NGC\,5850 has likely been perturbed by a
high-speed encounter with the nearby massive elliptical NGC\,5846.

\section{Molecular Gas and other Tracers}
\begin{figure}
\center{\resizebox{75mm}{75mm}{\includegraphics{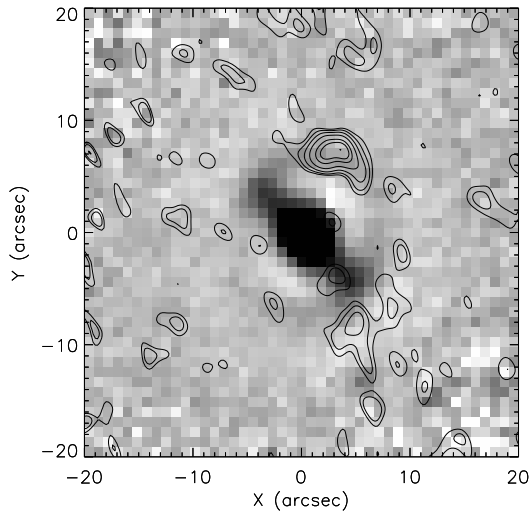}}}
\caption{Small-scale details (spatial filtering scale $< 1.6''$) on
  the J-image overlaid by CO(1$\to$0) integrated intensity contours (0.01,
  0.012, 0.016, 0.02, 0.025, 0.03, 0.04, 0.05, 0.063, 0.08, 0.1\,Jy/beam).}
\label{fig_co_imj}
\end{figure}

The CO(1-0) emission has been mapped in NGC\,5850, i) in the very
center, using the IRAM Plateau de Bure interferometer, to reach a
2.4\arcsec$\times$1.5\arcsec\ (PA$=$$-165^\circ$) spatial resolution
(Fig.~\ref{fig_co_imj}), and ii) in the primary bar with the IRAM-30m
telescope, with a 22\arcsec\ beam. We estimate the flux missed by the
interferometer to about 40\% of the single-dish flux. The total flux
in the center is 41.7\,Jy/beam for the interferometer.  To estimate
the H$_2$ surface density in the center, we used the standard
conversion factor N(H$_2$)/I$_{\rm CO}=2.3\cdot10^{20} \, \rm
cm^{-2}\,K^{-1}\,km^{-1}\,s$.  The northern concentration of gas
reaches a surface density of $200\,\rm M_\odot\,pc^{-2}$ (not
including 30\% He).  The total H$_2$ mass towards the center is about
$6.7\cdot10^7\,\rm M_\odot$ with the interferometer. Using the
single-dish, we find that the primary bar has about $1.5\cdot10^9\,\rm
M_\odot$ of molecular gas.

The critical surface density $\Sigma_c$ for gravitational
instabilities (e.g. Kennicutt 1989) is:
$\Sigma_c=\alpha\frac{\kappa\sigma}{3.36G}$, with $\kappa$ the
epicyclic frequency, $\sigma$ the velocity dispersion of the gas and
$\alpha$ is a constant close to unity.  The northern concentration,
likely made of a collection of Giant Molecular Clouds (GMCs), has a
{\em global} velocity dispersion of $\sim$100\,km/s.  The critical
surface density is then $900\,\rm M_\odot\,pc^{-2}$ which is much
higher than the peak value observed.
It may explain why no star formation, as traced by the H$\alpha$
emission, has occured in that great reservoir of molecular gas. The
H\,{\sc i} gas (total mass of $3.3\cdot10^9\,\rm M_\odot$) is more
concentrated in the larger inner ring and outer arms of the galaxy
(Higdon et al. 1998).


\section{Conclusion}
We have found CO emission in the center of NGC\,5850, located in a
single peak on the northern part of the nuclear ring.  The high
velocity dispersion of the molecular gas may prevent star formation in
that region. The CO distribution in the center of barred galaxies is
generally found to be either in the nucleus (Garc\'{\i}a-Burillo et
al.  1998) or trapped in twin peaks related to the resonances of the
bars (Kenney et al. 1992; Downes et al. 1996; Garc\'{\i}a-Burillo et
al.  1998).  Gas simulations performed with a single bar pattern and
without the tidal influence of the companion NGC\,5846 are unable to
reproduce the features observed in NGC\,5850 (Combes, Leon, Friedli \& Buta
1998, in preparation).  The decoupling of a second bar appears
necessary. The presence of the single molecular peak could be due to
an $m=1$ mode excited by the massive companion.

\small{

}
\end{document}